\documentclass[smallextended]{svjour3}       

\smartqed  
\usepackage{latexsym}
\usepackage{graphicx} 
\usepackage{amsfonts}

\usepackage{subfigure}
\usepackage{float}

\usepackage{amsmath} 

\usepackage{amssymb}
\usepackage{stmaryrd}
\usepackage{enumerate}
\usepackage{amsmath}
\usepackage{natbib}

\usepackage{subfigure}
\usepackage{tikz}
\usetikzlibrary{arrows}

\usepackage{slashbox}

\usepackage{bbm}
\usepackage{bm}
\usepackage{color}
\usepackage{flushend}
\usepackage{multirow}
\usepackage{makecell}
\usepackage{color}
\usepackage{colortbl}

\usepackage{amsfonts}
\usepackage{color}

\usepackage{amssymb}

\def\BibTeX{{\rm B\kern-.05em{\sc i\kern-.025em b}\kern-.08em
    T\kern-.1667em\lower.7ex\hbox{E}\kern-.125emX}}

\begin{document}

%
\title{Issues in the Multiple Try Metropolis mixing}
\author{L. Martino$^\star$ \and F. Louzada$^\star$}
\institute{$^\star$  Institute of Mathematical Sciences and Computing, \\
Universidade de S\~ao Paulo, S\~ao Carlos (S\~ao Paulo). \\
{\tt lukafree@icmc.usp.br}
}
\date{Received: date / Accepted: date}

\maketitle

\begin{abstract}
The multiple Try Metropolis (MTM) algorithm
is an advanced MCMC technique based on drawing and testing several candidates at each iteration of the algorithm. One of them is selected according to certain weights and then it is tested according to a suitable acceptance probability. Clearly, since the  computational cost increases as the employed number of tries grows, one expects that the performance of an MTM scheme improves as the number of tries increases, as well. However, there are scenarios where the increase of number of tries does not produce a corresponding enhancement of the performance. In this work, we describe  these scenarios and then we introduce possible solutions for solving these issues.
\end{abstract}

\section{Introduction}

Markov chain Monte Carlo (MCMC) methods are classical Monte Carlo techniques \citep{Robert04}, that produce a Markov chain converging to a target probability density function (pdf), usually to approximate an otherwise-incalculable integral \citep{Liu04b,Liang10}. 

The  Multiple-Try Metropolis (MTM) method \citep{Liu00} is an extension of the Metropolis-Hastings algorithm \citep{Metropolis53,Hastings70} in which the next state of the chain is selected among a set of $N$ independent and identically distributed (i.i.d.) samples. This enables the MTM sampler to make large step-size jumps without a lowering in the acceptance rate; and thus MTM  can explore easily a larger portion of the sample space in fewer iterations.  Different MTM schemes have been proposed in literature \citep[Chapter 13]{Frenkel96}, \citep{Qin01,Casarin2011,Pandolfi10,LucaJesse1,Craiu07} and have been studied in several works \citep{Bedard12,LucaJesse2,MartinoMTM2014}. More recently parallel MTM algorithms have been proposed in \citep{O-MCMC}. 

A well-designed MTM scheme improves its performance as the number of tries, $N$,  grows. Namely, when $N$ grows approaching infinity,  
 the correlation among the generated samples should vanish to zero. Clearly, this is at the expense of an increasing computational cost due to the use of a greater number of tries. In this work, we describe certain scenarios where the use of a greater $N$ in a standard MTM method \citep{Liu00}  and its extensions \citep{Casarin2011,Pandolfi10,LucaJesse1,LucaJesse2} does not yield an improvement in the performance. We explain the reasons of these drawbacks, and provide possible solutions for fixing these issues. The first scenario involves the use of a  single random-walk proposal within a standard MTM structure, whereas, in the second scenario, the use of multiple proposal pdfs independent from the previous state of the chains  is considered. In the first one, the increase of number of tries is always prejudicial, regardless of the choice of the weight functions (involving the target function in a suitable way \citep{Liu00,LucaJesse2}). In the second one, the increase of number of tries can help the mixing of the chain using a certain class of the weight functions (clearly, at the expense of a greater computational cost). However, we discuss different ways of using the set of multiple independent proposal pdfs within an MTM scheme improving the performance, in any case. For improving the performance in the first scenario, we suggest to use an MTM with variable number of tries, in a suitable way without jeopardizing the ergodicity of the chain.

\section{Multiple Try Metropolis with a single random-walk proposal}
\label{SectMPM}

\begin{table}[!t]
\caption{Multiple Try Metropolis with a (single) random-walk proposal (RW-MTM).}
\begin{tabular}{|p{0.95\columnwidth}|}
    \hline
\footnotesize
\begin{enumerate}
\item  Draw $N$ independent samples from the proposal pdf, 
$$
{\bf z}_1,\ldots,  {\bf z}_N\sim q({\bf x}|{\bf x}_{t-1})=q({\bf z}-{\bf x}_{t-1}).
$$
\item Select a sample ${\bf z}\in\{{\bf z}_1,\ldots,{\bf z}_N\}$, according to the probabilities 
\begin{equation}
\label{Eq_betas}
\bar{w}_k=\frac{w({\bf z}_k|{\bf x}_{t-1})}{\sum_{n=1}^N w({\bf z}_n|{\bf x}_{t-1})}, \quad \mbox{ where } \quad w({\bf z}_k|{\bf x}_{t-1})=\frac{\pi({\bf z}_k)}{q({\bf z}_k|{\bf x}_{t-1})},
\end{equation}
for $k=1,\ldots,N$.
\item Draw $N-1$  auxiliary points from the proposal $q$ given the previous selected sample ${\bf z}$, namely ${\bf y}_1,\ldots, {\bf y}_{N-1}\sim q({\bf x}|{\bf z})$, and set ${\bf y}_N={\bf x}_{t-1}$.
\item Compute the weights of the auxiliary points,
\begin{equation}
\label{Eq_betas2}
w({\bf y}_k|{\bf z})=\frac{\pi({\bf y}_k)}{q({\bf y}_k|{\bf z})}, \quad \mbox{ for } \quad k=1,\ldots,N.
\end{equation}
\item Set ${\bf x}_t= {\bf z}$ with probability
\begin{equation}
\label{AlfaMTM}
\alpha({\bf x}_{t-1},{\bf z})=\min\left[1,\frac{   \sum_{n=1}^N  w({\bf z}_n|{\bf x}_{t-1})   }{  \sum_{n=1}^N  w({\bf y}_n|{\bf z}) }  \right].
\end{equation}
Otherwise, set ${\bf x}_t= {\bf x}_{t-1}$, with probability $1-\alpha({\bf x}_{t-1},{\bf z})$. 
\end{enumerate} \\
\hline 
\end{tabular}
\label{MTM_Table}
\end{table}

Let us denote the target density as $\bar\pi({\bf x})\propto \pi({\bf x})$. First of all, we consider the use of a single random-walk proposal density, $q({\bf z}|{\bf x}_{t-1})=q({\bf z}-{\bf x}_{t-1})$. 
 Given a current state of the chain ${\bf x}_{t-1}\in \mathcal{X}\subseteq \mathbb{R}^{d_X}$, $t\in \mathbb{N}$, an MTM scheme generates $N$ independent candidates $\{{\bf z}_1,\ldots,{\bf z}_N\}$ from a proposal density $q$, i.e.,
\begin{gather}
\begin{split}
\nonumber
{\bf z}_1,\ldots, {\bf z}_N\sim q({\bf z}|{\bf x}_{t-1}).
\end{split}
\end{gather}
Then, one sample ${\bf z}$ is selected among the set $\{{\bf z}_1,\ldots, {\bf z}_N\}$, according to certain weight functions \citep{Liu00,LucaJesse2}. The movement from ${\bf x}_t$ to ${\bf z}$ is accepted with a suitable probability $\alpha({\bf x}_{t-1},{\bf z})$, which also depends on the rest  of candidates. The probability $\alpha({\bf x}_{t-1},{\bf z})$ is designed such that the kernel of the MTM algorithm  fulfills the detailed balance condition. Only for facilitating the comprehension, we consider the importance weights 
\begin{equation}
w({\bf z}_k|{\bf x}_{t-1})=\frac{\pi({\bf z}_k)}{q({\bf z}_k|{\bf x}_{t-1})}, 
\end{equation}
for choosing ${\bf z}\in \{{\bf z}_1,\ldots, {\bf z}_N\}$, i.e., ${\bf z}$ is selected according the probabilities $\bar{w}_k=\frac{w({\bf z}_k|{\bf x}_{t-1})}{\sum_{n=1}^N w({\bf z}_n|{\bf x}_{t-1})}$. Different kind of weights could be used \citep{LucaJesse2,Pandolfi10}, but without avoiding the problem that we describe in the next section. 

Table \ref{MTM_Table} shows all the details of the MTM technique. Observe that, an RW-MTM method requires the generation of $N-1$ auxiliary points ${\bf y}_1,\ldots,{\bf y}_{N-1}$ from $q(\cdot|{\bf z})$ (see Step 3 of Table \ref{MTM_Table}). Moreover, note that the selected sample ${\bf z}$ is drawn from the empirical measure
\begin{equation}
{\hat \pi}^{(N)}({\bf z})=\sum_{n=1}^N {\bar w}_n\delta({\bf z}-{\bf z}_n),
\end{equation}
that  approximates the distribution of $\pi$, via importance sampling (IS) \citep{Robert04,Liu04b}. Finally, we remark that the acceptance probability $\alpha({\bf x}_{t-1},{\bf z})$ in Eq. \eqref{AlfaMTM} can be expressed as
\begin{equation}
\alpha({\bf x}_{t-1},{\bf z})=\min\left[1,\frac{ \hat{Z}({\bf z}_1,\ldots,{\bf z}_N|{\bf x}_{t-1}) }{ \hat{Z}({\bf y}_1,\ldots,{\bf y}_N|{\bf z}) }  \right],
\end{equation}
where the function $\hat{Z}(\cdot|{\bf r}):  \mathcal{X}^N \rightarrow \mathbb{R}$, with ${\bf r}\in \mathcal{X}$,
\begin{equation}
\hat{Z}({\bf v}_1,\ldots,{\bf v}_N|{\bf r})=\frac{1}{N} \sum_{n=1}^N\frac{\pi({\bf v}_n)}{q({\bf v}_n|{\bf r})},
\end{equation}
 is an estimator of the normalizing constant $Z=\int_{\mathcal{X}} \pi({\bf x}) d {\bf x}$ \citep{Robert04}, i.e., of the area below $\pi({\bf x})$.



\section{Problem in the RW-MTM mixing}
\label{RW-MTMproblem}
The desired behavior of an MTM scheme is that the performance improves as the number of used candidates $N$ grows (jointly with the computational cost). Indeed in general, as $N$ increases, the chosen point ${\bf z}$ is selected from a better IS approximation ${\hat \pi}^{(N)}$ of $\bar{\pi}$, so that ${\bf z}$ is a better candidate to be tested as new possible state of the chain. As a consequence, in a well-designed MTM scheme the acceptance probability $\alpha({\bf x}_{t-1},{\bf z})$ should approach $1$ when $N\rightarrow \infty$. Thus, in general, MTM fosters greater ``jumps'' and, as a consequence, a faster exploration of the state space.  
However, below we describe a scenario where the increase of number $N$ of tries could be even damaging.

\begin{figure}[htb]
\centerline{
 	\includegraphics[width=8cm]{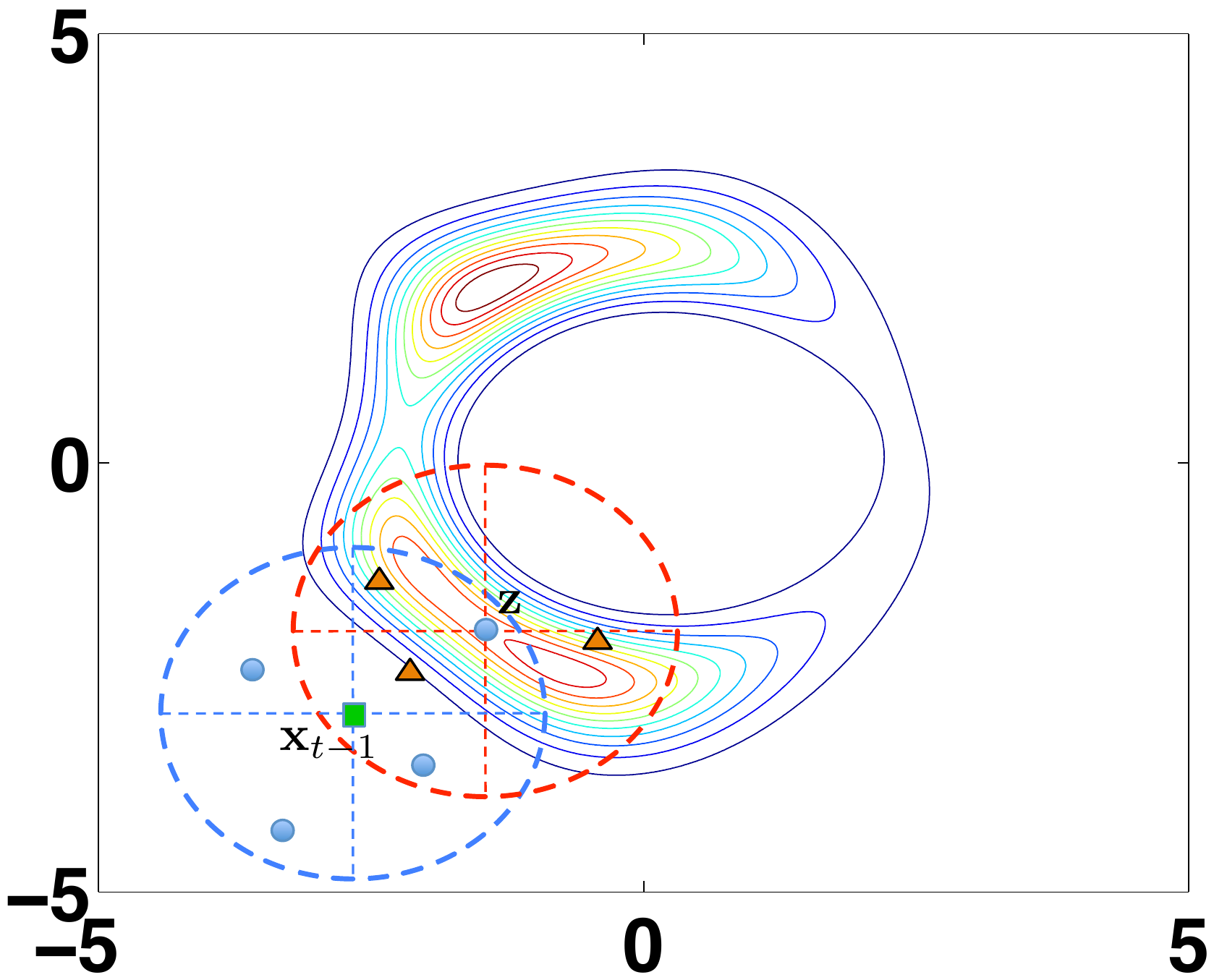}
 		}
\caption{Graphical representation of a possible scenario described in Section \ref{RW-MTMproblem}, where $\hat{Z}_2>\hat{Z}_1$  (and $\hat{Z}_2>>\hat{Z}_1$ when $N$ grows). We show the contour plot of a bidimensional target pdf $\pi({\bf x})$ with solid lines. The previous state of the chain ${\bf x}_{t-1}$ is depicted with a square; the $N=4$ candidates ${\bf z}_j$'s are shown with circles, whereas the $N-1=3$ auxiliary points ${\bf y}_i$'s are illustrated with triangles. Dashed lines represent the scale parameters of the proposal  densities $q(\cdot|{\bf x}_{t-1})$ and $q(\cdot|{\bf z})$, where ${\bf z}\in \{{\bf z}_1,\ldots,{\bf z}_4\}$ is the selected candidate. }
\label{Fig1}
\end{figure}

For facilitating the explanation, we assume that the expected value of the random variable ${\bf Z} \sim q({\bf z}-{\bf x}_{t-1})$ is exactly ${\bf x}_{t-1}$, i.e.,  $E[{\bf Z}]={\bf x}_{t-1}$, e.g., when $q$ is Gaussian, $q({\bf z}-{\bf x}_{t-1})=\mathcal{N}({\bf z};{\bf x}_{t-1},{\bf C})$. Let us denote $\hat{Z}_1=\hat{Z}({\bf z}_1,\ldots,{\bf z}_N|{\bf x}_{t-1})$ and $\hat{Z}_2=\hat{Z}({\bf y}_1,\ldots,{\bf y}_N|{\bf z})$, so that we can rewrite the acceptance probability as
\begin{equation}
\alpha=\min\left[1,\frac{ \hat{Z}_1 }{ \hat{Z}_2 }  \right].
\end{equation}
Furthermore, consider a scenario where the state in the $(t-1)$-th iteration, ${\bf x}_{t-1}$, is placed in a region of low probability of ${\bar \pi}({\bf x})\propto \pi({\bf x})$, nearby a region of high probability mass (e.g., see Figure \ref{Fig1}(a)). Assume also that the variance of the proposal $q({\bf z}-{\bf x}_{t-1})$ is wide enough in order to (at least) reach the region of high probability mass of $\pi$. In this situation,  several drawn tries are located in the region of small probability around the value $E[{\bf Z}]={\bf x}_{t-1}$. On the other hand, it is possible that few of them are located close to the mode of $\pi$;  Figure \ref{Fig1}(a) depicts a possible scenario of this kind, with only $N=4$ tries and one of them located in a mode of $\pi$.  Thus, it is highly probable that the MTM selected one well-located point as proposed sample ${\bf z}$, after the resampling at Step 2. For the same reasons, in general, many of the $N-1$ auxiliary points, ${\bf y}_1,\ldots,{\bf y}_{N-1}$ drawn from $q({\bf y}|{\bf z})$, will be placed around the mode of $\pi$. Hence, in this situation, we have that 
$$
\hat{Z}_2=\frac{1}{N} \sum_{n=1}^N\frac{\pi({\bf y}_n)}{q({\bf y}_n|{\bf z})} >>\hat{Z}_1=\frac{1}{N} \sum_{n=1}^N\frac{\pi({\bf z}_n)}{q({\bf z}_n|{\bf x}_{t-1})}.
$$ 
 As a consequence, 
$$
\alpha({\bf x}_{t-1},{\bf z})\approx 0,
$$ 
so that the chain can remain stuck at ${\bf x}_{t-1}$. It is important to observe that this situation can become even worse if $N$ grows. On the contrary, in this scenario, the use of a smaller number of tries can help to jump to the region of high probability. Finally, we remark that the problem previously described cannot be solved by changing of analytical form of the weights \citep{Liu00,LucaJesse2}.\footnote{A suitable acceptance function $\alpha$ for generic weight functions is shown in Appendix \ref{NoSol}, for the case of multiple independent proposal densities.  }

\subsection{Proposed solution}
\label{PropSol1}

Let us denote as $K_m({\bf x}_t|{\bf x}_{t-1},N_m)$ the kernel of an MTM scheme employing $N_m$ tries. We consider  a combination $M$ different kernels each of which using a different number of tries $N_m$, $m=1,\ldots,M$, i.e.,
\begin{equation}
\label{Ducati}
K({\bf x}_t|{\bf x}_{t-1})=\frac{1}{M} \sum_{m=1}^M K_m({\bf x}_t|{\bf x}_{t-1},N_m).
\end{equation}
It is straightforward to show that if each  $K_m({\bf x}_t|{\bf x}_{t-1},N_m)$ leaves invariant $\pi$, also $K({\bf x}_t|{\bf x}_{t-1})$ has $\pi$ as invariant pdf \citep{Robert04,Liu04b}.  
Therefore, fixing the averaged computational effort, represented by the averaged number or tries 
$$
\widetilde{N}=\frac{1}{M}\sum_{m=1}^M N_m,
$$
we choose $M$ different values $N_m\in \mathbb{N}$, such that $\widetilde{N}$ is the desired one.
The idea is to use a variable number of tries, i.e., a different number of candidates at each iteration. Namely, at each iteration, an index $m'$ is drawn uniformly within $1,\ldots,M$ and then $N_{m'}$ tries are employed in the MTM scheme $K_{m'}$. Note that this is equivalent to use the kernel in Eq. \eqref{Ducati}. Choosing at least one small value, e.g., $N_1=1$, this helps jumps of the chain in the awkward scenario, previously described. See the numerical simulations for further details.

\section{Multiple Try Metropolis with different independent proposals}
The MTM algorithm  in Table \ref{MTM_Table} can be simplified if the proposal pdf $q({\bf x})$ is independent from the previous state of the generated chain. Indeed, in this case, Step 3 in Table \ref{MTM_Table} can be removed, in the sense that it is possible to avoid the generation of the auxiliary points \citep{Liu00,LucaJesse2}. Furthermore, it is also possible to employ simultaneously different proposal pdfs $q_1({\bf x}),\ldots, q_N({\bf x})$ \citep{Casarin2011,LucaJesse2}. The resulting algorithm is detailed in Table \ref{MTM_Table2}, considering the use of importance weights. 
The acceptance probability $\alpha$ in Eq. \eqref{AlfaMTM2} can be written again as 
$$
\alpha=\min\left[1,\frac{ \hat{Z}_1 }{ \hat{Z}_2 }  \right],
$$
where, in this case,
\begin{eqnarray}
\label{Zeq}
\hat{Z}_1&=& \frac{1}{N}\sum_{n=1}^N w_n({\bf z}_n), \nonumber\\
\hat{Z}_2&=&\frac{1}{N}\left(N \hat{Z}_1-w_j({\bf z}_j)+w_j({\bf x}_{t-1})\right).
\end{eqnarray}
The general acceptance function $\alpha$ for I-MTM using generic (bounded and positive) weights is shown in Eq. \eqref{Estoca}.
\begin{table}[!t]
\caption{Multiple Try Metropolis with different independent proposals (I-MTM).}
\begin{tabular}{|p{0.95\columnwidth}|}
    \hline
\footnotesize
\begin{enumerate}
\item  Draw $N$ independent samples 
$$
{\bf z}_1\sim q_1({\bf x}),\ldots,  {\bf z}_N\sim q_N({\bf x}).
$$
\item Select a sample ${\bf z}_j\in\{{\bf z}_1,\ldots,{\bf z}_N\}$, according to the probabilities 
\begin{equation}
\label{Eq_betas1}
\bar{w}_k=\frac{w_k({\bf z}_k)}{\sum_{n=1}^N w_n({\bf z}_n)}, \quad \mbox{ where } \quad w_k({\bf z}_k)=\frac{\pi({\bf z}_k)}{q_k({\bf z}_k)},
\end{equation}
for $k=1,\ldots,N$.
\item Set ${\bf x}_t= {\bf z}_j$ with probability
\begin{eqnarray}
\label{AlfaMTM2}
\alpha({\bf x}_{t-1},{\bf z}_j) 
&=&\min\left[1,\frac{   \sum_{n=1}^N  w_n({\bf z}_n)   }{  \sum_{n=1}^N  w_n({\bf z}_n)-w_j({\bf z}_j)+w_j({\bf x}_{t-1}) }  \right]. 
\end{eqnarray}
Otherwise, set ${\bf x}_t= {\bf x}_{t-1}$, with probability $1-\alpha({\bf x}_{t-1},{\bf z}_j)$. 
\end{enumerate} \\
\hline 
\end{tabular}
\label{MTM_Table2}
\end{table}

\section{Problem in the I-MTM mixing}
\label{OtroProbl}
First of all, we can observe that the sums in $\hat{Z}_1$ and $\hat{Z}_2$ in Eq. \eqref{Zeq} differ only for one weight, i.e., $\hat{Z}_1$ contains $w_j({\bf z}_j)$ but does not involve $w_j({\bf x}_{t-1})$, whereas $\hat{Z}_2$ includes  $w_j({\bf x}_{t-1})$, instead of $w_j({\bf z}_j)$. Thus, using importance weights, the probability $\alpha$ of an I-MTM scheme always approaches $1$ when $N$ increases, if the employed weight functions are included in the class of weights proposed in \citep{Liu00}.\footnote{{\color{black} Considering the case of independent proposal pdfs, the class of weights in \citep{Liu00} is defined as $w_k({\bf y}_k|{\bf {\bf z}})=\pi({\bf z}_k)q_k( {\bf x})\lambda_k({\bf z}_k, {\bf x})$ with $k=1,\ldots, N$, and $\lambda_k({\bf z}_k, {\bf x})=\lambda_k({\bf x},{\bf z}_k)$ is a generic symmetric function w.r.t. ${\bf z}_k$ and ${\bf x}$. As an example, if we set $\lambda_k({\bf z}_k, {\bf x})=\frac{1}{q_k( {\bf x})q_k({\bf z}_k)}$, we obtain the importance weights $w_k({\bf z}_k|{\bf {\bf x}})=w_k({\bf z}_k)=\frac{\pi({\bf z}_k)}{q_k({\bf z}_k)}$.}
 } 
This statement is instead not valid, in general, for the generic weight functions given in \citep{Pandolfi10,LucaJesse2} and recalled in Eq. \eqref{Estoca}.
 
\begin{figure}[htb]
\centerline{
 	\includegraphics[width=8cm]{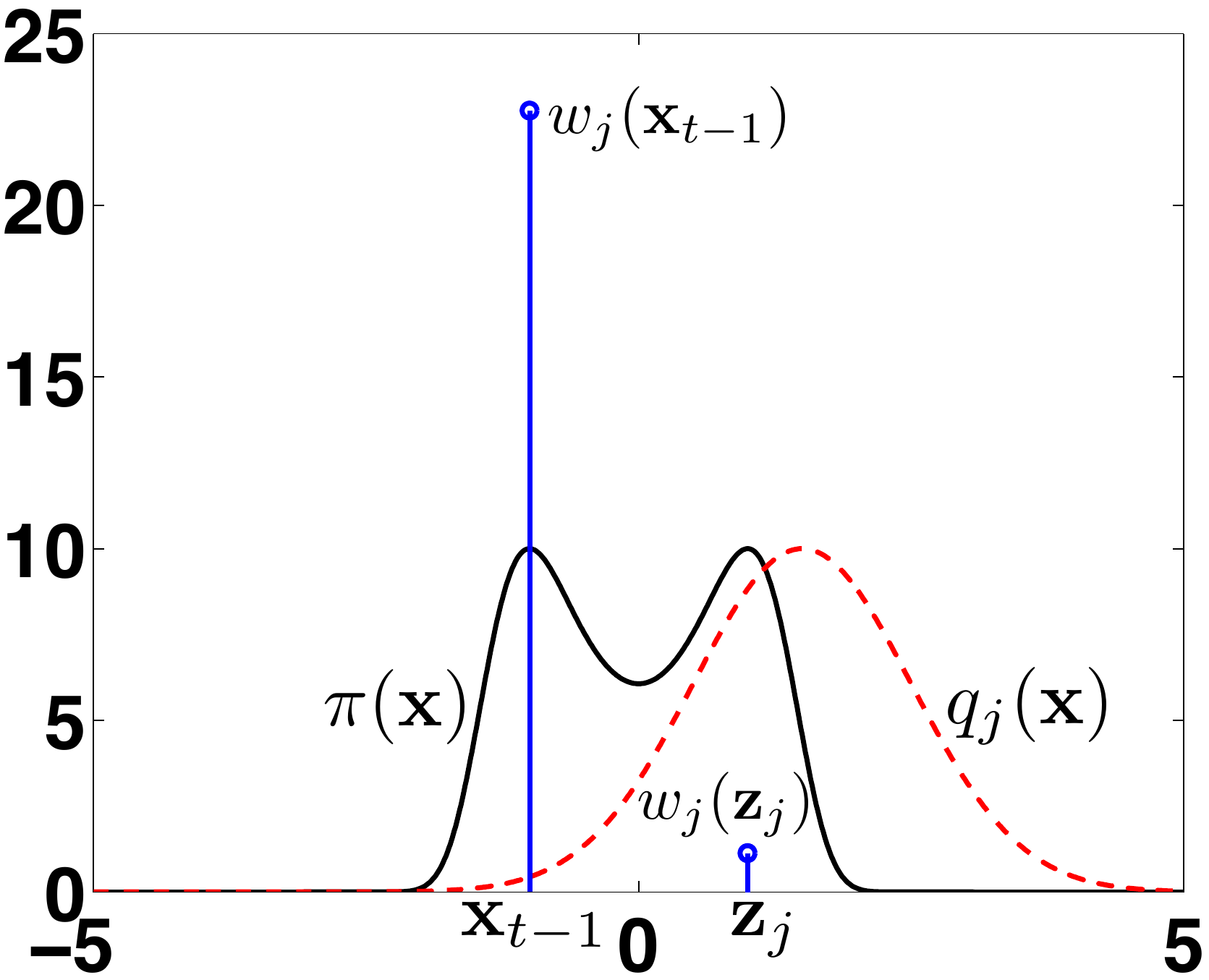}
 		}
\caption{Graphical representation of the scenario described in Section \ref{OtroProbl}. The contour plot of a bimodal (unnormalized) target pdf $\pi({\bf x})$ is depicted with solid line whereas the $j$-th (unnormalized) proposal pdf $q_j({\bf x})$ is shown with dashed line.
 }
\label{Fig2}
\end{figure}

  In this section we focus on the use of importance weights, which are contained in class discussed in \citep{Liu00}. The solutions that we discuss later on are valid in any cases, including the use of the generic weights in Appendix \ref{NoSol}. Note that, in I-MTM, the $j$-th weight involves the $j$-th proposal pdf, i.e.,
$$
w_j({\bf x})=\frac{\pi({\bf x})}{q_j({\bf x})}.
$$
We need to evaluate the $j$-th weight $w_j$, involving the $j$-th proposal $q_j$, at ${\bf z}_j$ and ${\bf x}_{t-1}$. The sample ${\bf z}_j$ is drawn from $q_j$ by definition, whereas ${\bf x}_{t-1}$ is the previous state of the chain (it could be generated from any possible $q_n$ in the previous iterations of the I-MTM algorithm). Hence, with high probability  ${\bf z}_j$ is located nearby a mode of $q_j$, since $ {\bf z}_j\sim q_j({\bf z})$, whereas ${\bf x}_{t-1}$ could be placed close to a mode or a tail of $q_j$ with equal chance, in general. Thus, since the proposal $q_j$ appears in the denominator of the weights $w_j$, in general we have $w_j({\bf z}_j) < w_j({\bf x}_{t-1})$, producing small values of acceptance probability $\alpha$, if $N$ is not enough big. This scenario becomes even more complicated, if  the proposal pdf $q_j$ is placed close to a mode of the target $\pi$, and the previous state ${\bf x}_{t-1}$ is located in a tail of $q_j$. In this case, if $\pi({\bf x}_{t-1})\neq 0$, the value of $w_j({\bf x}_{t-1})$ can be huge and $w_j({\bf x}_{t-1})>> w_j({\bf z}_j)$. Hence, the I-MTM scheme tends to select several times the sample drawn from $q_j$, i.e., ${\bf z}_j$, as ``good'' candidate (step 2 of Table \ref{MTM_Table2}), but the movement from ${\bf x}_{t-1}$ to ${\bf z}_j$ is often rejected since $\alpha\approx 0$. As a consequence, the chain can remain indefinitely trapped in this situation.  Figure \ref{Fig2} represents graphical sketch  of this situation.

\subsection{Proposed solutions}
\label{propSol2}

Below, we discuss  different possible solutions, ordered for increasing theoretical complexity and practical interest. It is important to remark that the change of the analytic form of the weights is not a solution as shown in \ref{NoSol}.
\newline
{\bf First solution.}  First of all, let us consider the possibility of using a greater number of tries keeping fixed the number $N$ of proposal pdfs, i.e., denoting with $P$ the number of  tries we have $P>N$ with $P=kN$ with $k\in \mathbb{N}$. The problem described above could be solved increasing $P$, when the used weights are importance weights.{\color{black}\footnote{{\color{black} When other kind of weights is employed, the problem could persist even increasing $P$.}}} If ${\bf x}_{t-1}$ is located in a tail of $q_j$, the value of $P$ required to solve the issue, could be huge.
However, this trivial solution entails an increase of the computational cost in terms of evaluations of the target function. In the sequel, we introduce alternative solutions which do not require to increase the computational cost and are valid for any possible kind of weight functions, used within I-MTM.  
\newline
{\bf Second solution.} The problem described above disappears if we consider a unique proposal pdf defined as mixture, i.e.,
$$
\psi({\bf x})=\frac{1}{N} \sum_{n=1}^N q_n({\bf x}).
$$
Hence,  in this case, we draw ${\bf z}_1,\ldots,{\bf z}_N$ from $\psi({\bf x})$  and the weights  are 
$$w({\bf z}_n)=\frac{\pi({\bf z}_n)}{\psi({\bf z}_n)}.$$ 
We can observe that in the denominator of the importance weight all the components $q_n$'s are used and hence evaluated, in this case. Let us assume that the previous state of the chain ${\bf x}_{t-1}$ was generated from the $k$-th component of the mixture, i.e.,  $q_k({\bf x})$, in a previous iteration, and the selected candidate $z_j$ has been drawn from $q_j({\bf x})$, by definition. In this scenario, both pdfs, $q_k$ and $q_j$, are involved simultaneously in the denominator of importance weights, avoiding the problem previously described. Although the mixture $\psi({\bf x})$ takes into account all the proposal pdfs $q_n$'s, unlike in the I-MTM in Table \ref{MTM_Table2}, in this case only a subset of the components $\{q_1({\bf x}),\ldots,q_N({\bf x})\}$ participates in generating candidates at each iteration. To avoid this drawback, see below the next solution.
\newline
{\bf Third solution.}  The joint use of the functions $q_1({\bf x}),\ldots,q_N({\bf x})$ (with equal proportion, at each iteration) in general increases the robustness of the resulting algorithm. Namely, if no information is available to choose the best proposal in the set $\{q_1({\bf x}),\ldots,q_N({\bf x})\}$, a more robust strategy consists in employing always the complete set of functions. The {\it deterministic mixture} (DM) approach \citep{Veach95,Owen00,LetterVictor,TutorialMIS},  successfully applied in different sophisticated Monte Carlo algorithms \citep{CORNUET12,APIS15,LAIS2015}, provides a possible solution. Indeed, using the DM approach, we can draw one sample ${\bf z}_n$ from each proposal pdf $q_n({\bf x})$, i.e.,
$$
{\bf z}_1\sim q_1({\bf x}),\ldots,  {\bf z}_N\sim q_N({\bf x}),
$$ 
exactly as in step 1 of Table \ref{MTM_Table2}, and then assign the corresponding DM weights
$$
w({\bf z}_n)=\frac{\pi({\bf z}_n)}{\psi({\bf z}_n)}=\frac{\pi({\bf z}_n)}{\frac{1}{N} \sum_{n=1}^N q_n({\bf x})}, \quad n=1,\ldots,N. 
$$
It is possible to show that this approach is valid and it can be interpreted as variance reduction technique for sampling from a mixture of pdfs. Namely, we use a quasi-Monte Carlo approach for generating the indices $j_n$, $n=1,\ldots,N$, i.e., the deterministic sequence $j_1=1,j_2=2,\ldots,j_n=N$, and then ${\bf z}_n \sim  p({\bf x}|j_n)=q_n({\bf x})$ for $n=1,\ldots,N$. The DM approach improves the performance of the IS numerical approximation \citep{Owen00,LetterVictor}. Observe that, also in this case,  we solve the issue, since again all the proposals are included in the denominator of the weights, and we always use all the proposals $q_1,\ldots,q_N$ at each iteration (as in Table \ref{MTM_Table2}).



\section{Numerical simulations: localization in a wireless sensor network}
\label{sec:gibbsToy2}

We consider the problem of positioning a target $\textbf{X}$ in a two-dimensional space using range measurements  \citep{Ali07,Fitzgerald01}.
More formally, we consider a random vector $\textbf{X}=[X_1,X_2]^{\top}$ denoting the target's position in $\mathbb{R}^{2}$.
The measurements are obtained from $6$ sensors located at $\textbf{h}_1=[-5,1]^{\top}$, $\textbf{h}_2=[-2,6]^{\top}$, $\textbf{h}_3=[0,0]^{\top}$,  $\textbf{h}_4=[5,-6]^{\top}$, $\textbf{h}_5=[6,4]^{\top}$ and $\textbf{h}_6=[-4,-4]^{\top}$, and 
the observation equations are given by
\begin{gather}
\label{sistemaejemplo}
\begin{split}
R_{j} = -10\log\left(\frac{||{\bf X}-{\bf h}_j||}{0.3}\right) + \Omega_{j}, \quad j=1,\ldots, 6, \\
\end{split}
\end{gather}
where $\Omega_{j}$ are i.i.d. Gaussian random variables, $\Omega_j \sim \mathcal{N}(\omega_j;0,5)$.
Let us assume to receive the observation vector $\textbf{r} = [26, 26.5, 25,28,28, 25.3]^{\top}$.
In order to perform Bayesian inference, we consider a non-informative prior over ${\bf X}$ (i.e., an improper uniform density on $\mathbb{R}^{2}$), and study the posterior pdf, $\bar{\pi}({\bf x})=p(\textbf{x}|\textbf{r})\propto p(\textbf{r}|\textbf{x})p(\textbf{x})$. A contour plot of $\bar{\pi}({\bf x}) \propto \pi({\bf x})$ is shown in Figure \ref{Fig1}.

\begin{figure}[htb]
\centerline{
\subfigure[Stand. RW-MTM]{\includegraphics[width=4.5cm]{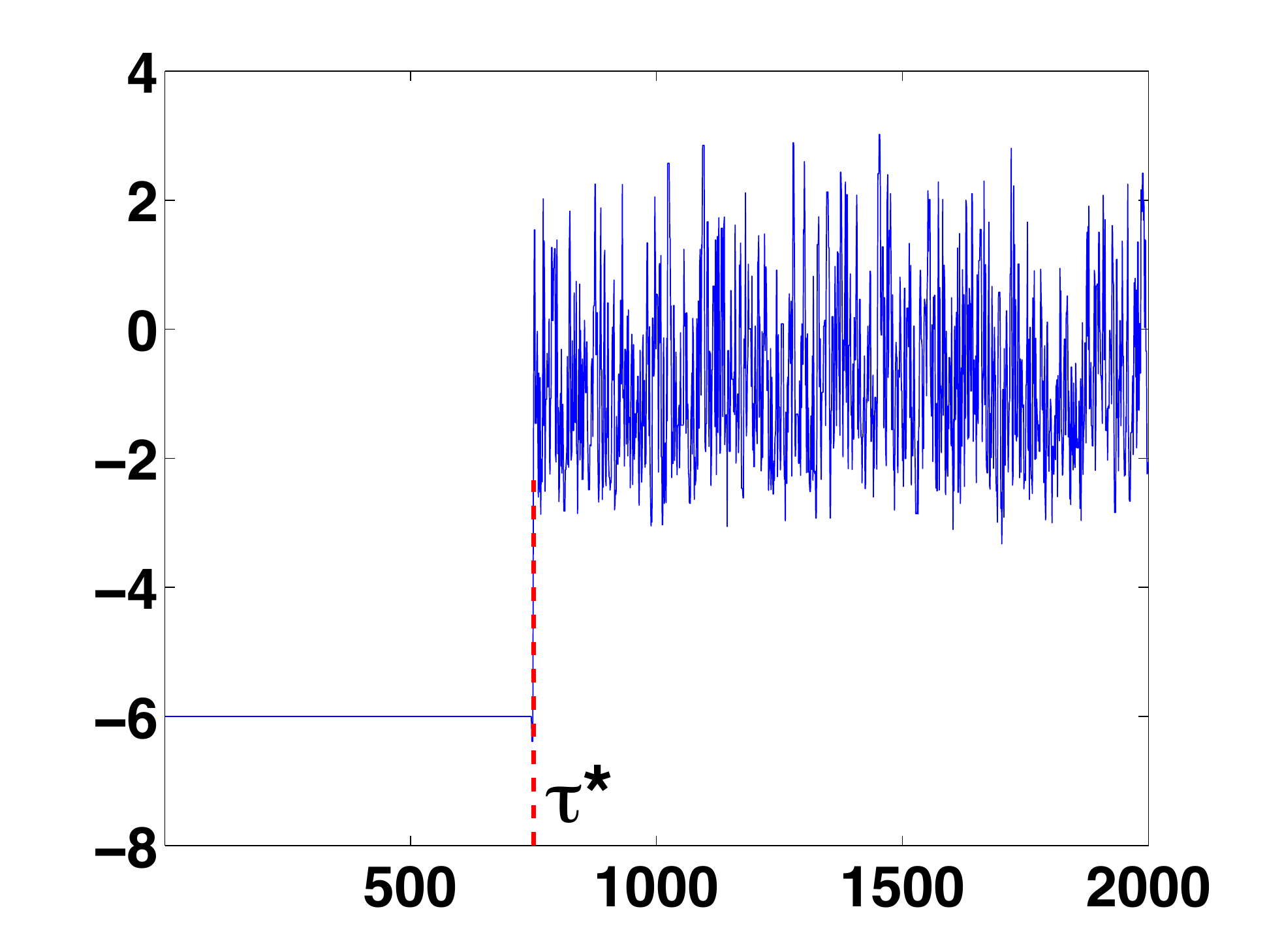}}
\subfigure[Stand. RW-MTM]{\includegraphics[width=4.5cm]{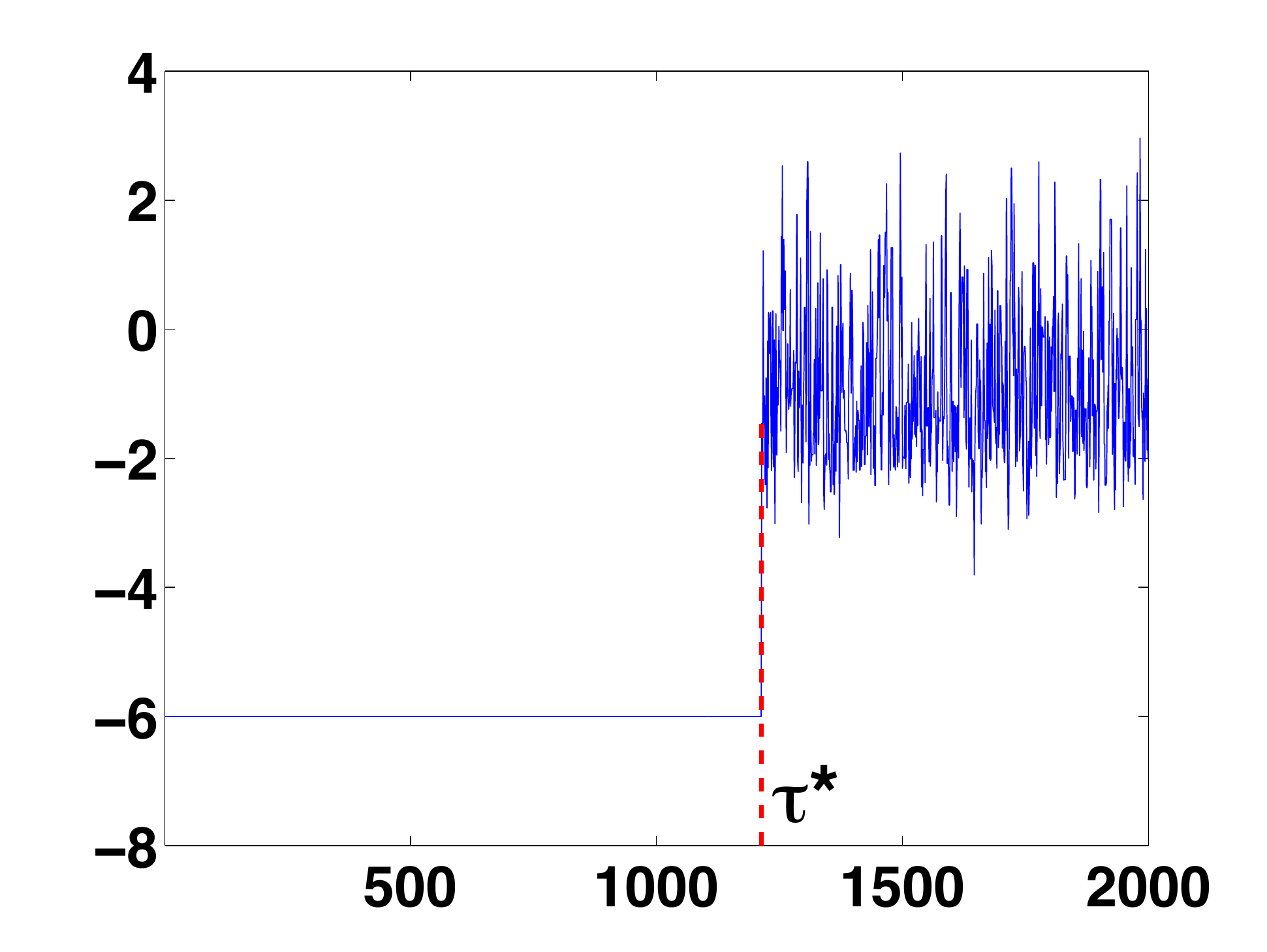}}
 	\subfigure[Stand. RW-MTM]{\includegraphics[width=4.5cm]{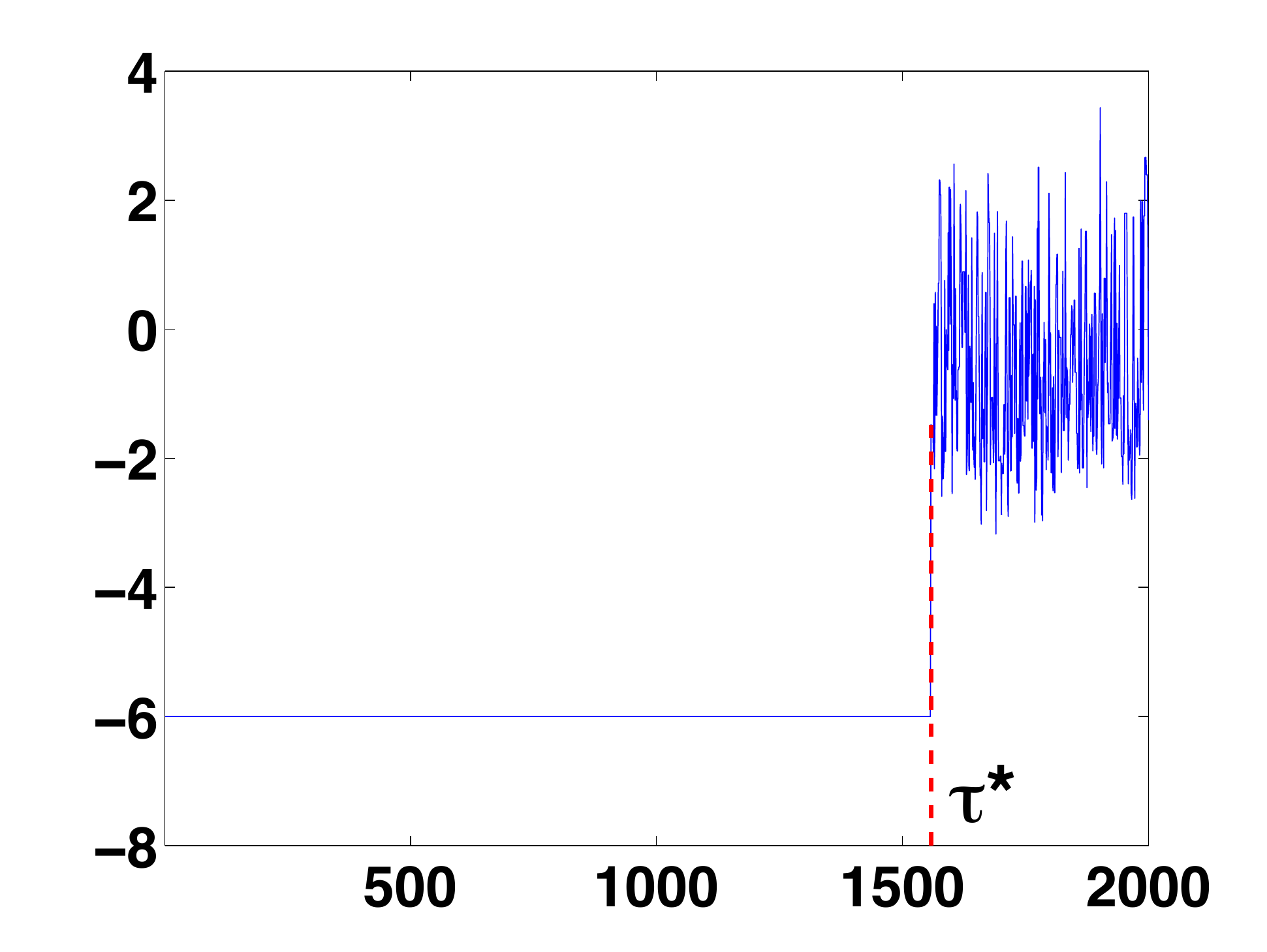}}
 		}
		\centerline{
\subfigure[Novel scheme]{\includegraphics[width=4.5cm]{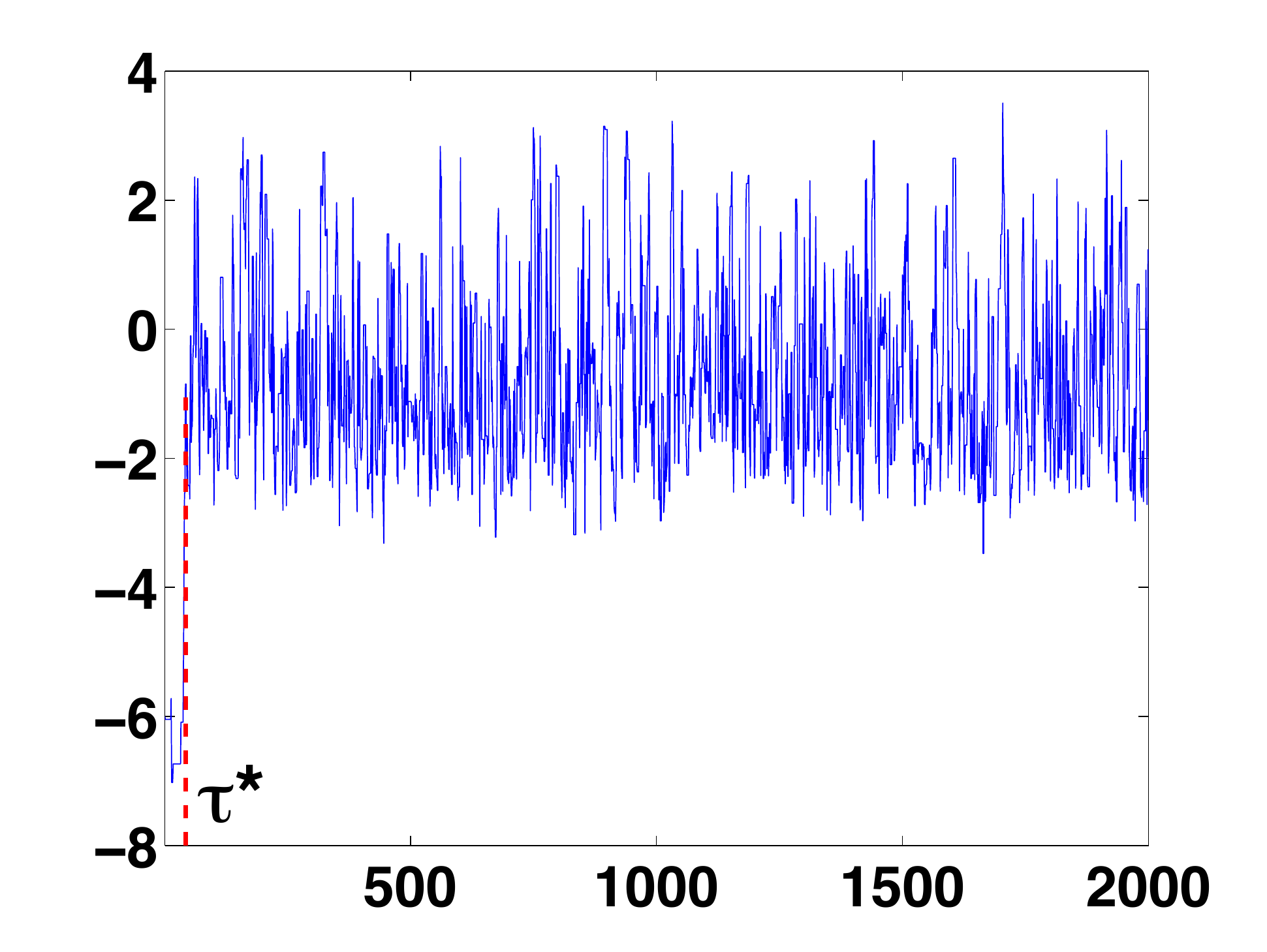}}
\subfigure[Novel scheme]{\includegraphics[width=4.5cm]{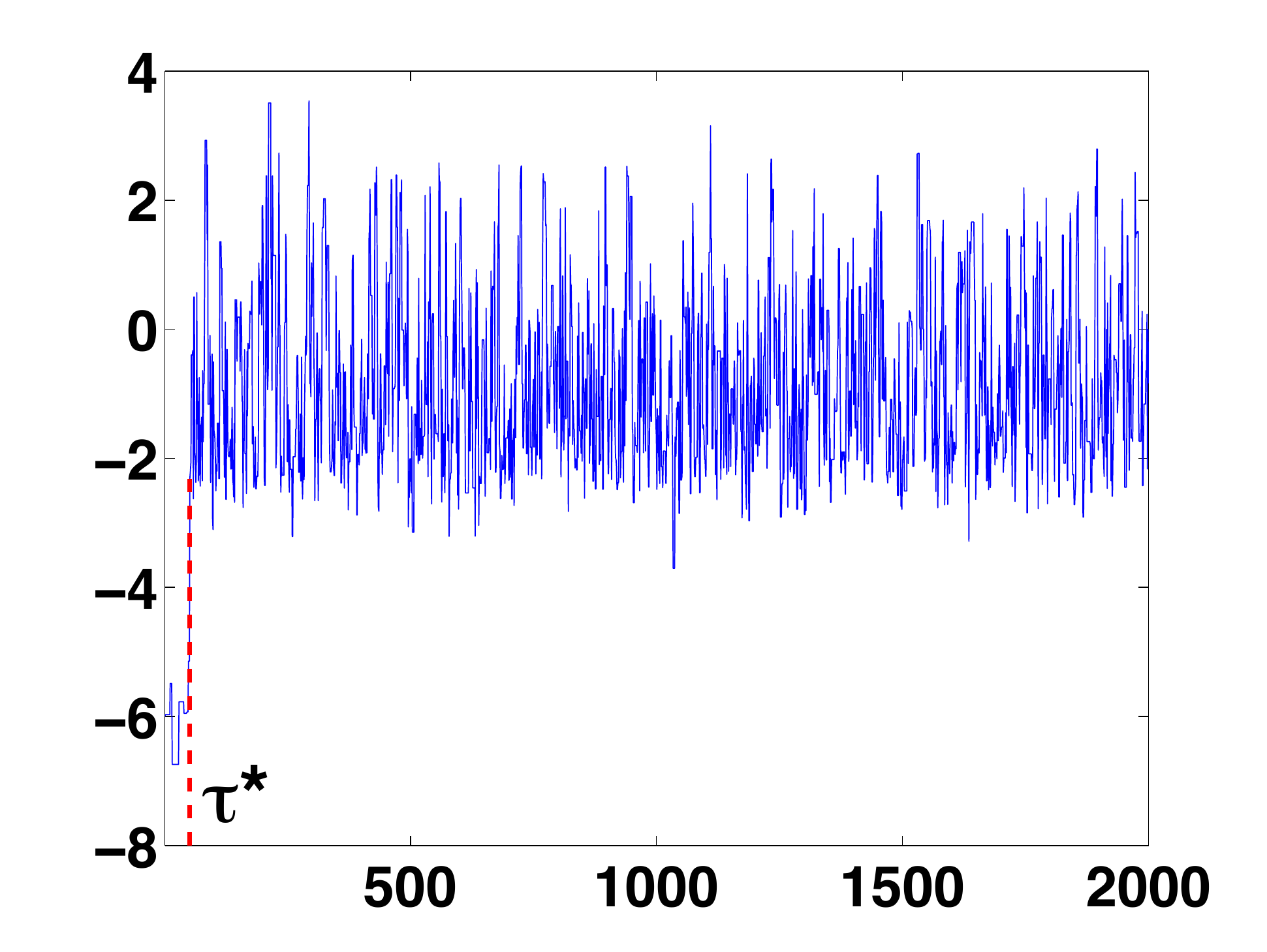}}
 	\subfigure[Novel scheme]{\includegraphics[width=4.5cm]{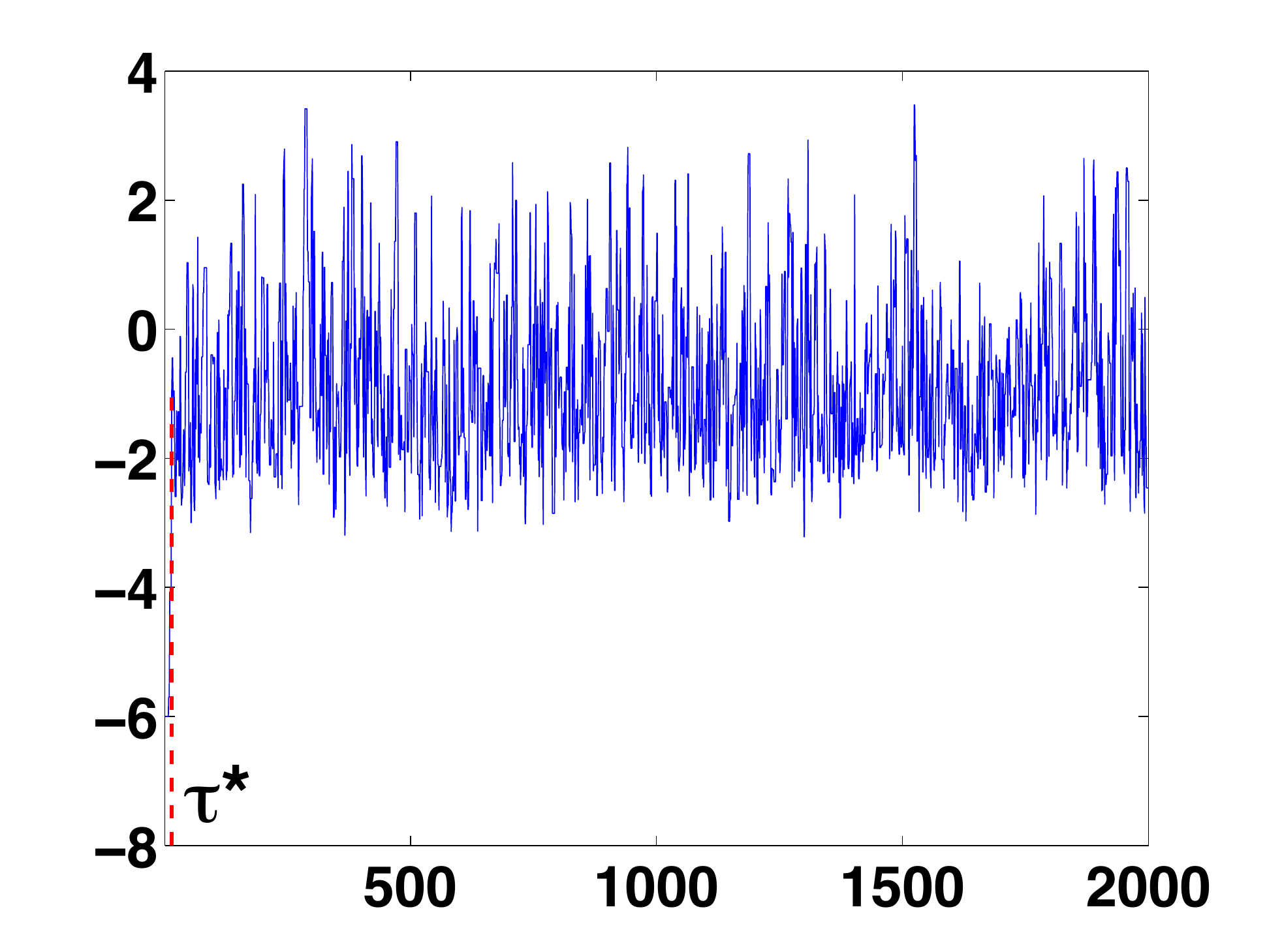}}
 		}
\caption{{\bf (a)}-{\bf (b)}-{\bf (c)} Realizations of the standard RW-MTM method with {\bf (a)} $N=\widetilde{N}=200$ ($\tau^*=750$, in this specific run), {\bf (a)} $N=\widetilde{N}=500$ ($\tau^*=1214$) and {\bf (c)} $N=\widetilde{N}=1000$ ($\tau^*=1558$). {\bf (d)}-{\bf (e)}-{\bf (f)} Realizations of the novel method with {\bf (a)} $\widetilde{N}=200$ ($\tau^*=43$, in this run), {\bf (a)} $\widetilde{N}=500$ ($\tau^*=52$) and {\bf (c)} $\widetilde{N}=1000$ ($\tau^*=15$).}
\label{FigSimu}
\end{figure}

We perform different MTM schemes for drawing samples from the posterior $\bar{\pi}({\bf x})$.  In order to highlight  the described issues, we decide the starting point of the chain at ${\bf x}_0=[-6,-6]^{\top}$ forcing the chain to escape from a region of low probability of $\bar{\pi}({\bf x})$. We run $500$ independent simulations of different MTM schemes with $t=1,\dots, T$ (we set $T=2000$ for RW-MTM and $T=4000$ for I-MTM), and compute the expected time needed for the chain to escape from the region around ${\bf x}_0$ and reach the region containing the modes of the target.  For this purpose,  at each iteration of the algorithm, we calculate the Euclidean distances $d_{1,t}=||{\bf x}_t-{\bf x}_0||$ and $d_{2,t}=||{\bf x}_t-{\bm \mu}||$ where ${\bm \mu}=E_{\pi}[{\bf X}]=[-0.753,-0.037]^{\top}$ is the expected value of ${\bf X} \sim \bar{\pi}({\bf x})$.\footnote{We have computed the vector $E_{\pi}[{\bf X}]$ numerically, using a computational expensive thin grid in $\mathbb{R}^2$.} At each run, we obtain the first iteration $\tau^*$ such that $d_{1,\tau^*} >d_{2,\tau^*}$, hence $\tau^*$ can be interpreted as the time that the chain remained trapped around ${\bf x}_0$, in the specific run ( see Figures \ref{FigSimu} as examples of $\tau^*$). Cleary, we have $1\leq \tau^* \leq T$. 
We repeat the procedure for $500$ independent runs, in order to approximate the expected time $E[\tau^*]$.  
\newline
{\bf RW-MTM.} For the random walk MTM method, we consider a Gaussian proposal $q({\bf x}|{\bf x}_{t-1})=\mathcal{N}({\bf x};{\bf x}_{t-1},{\bm \Sigma})$ where ${\bm \Sigma}=\sigma^2 \mathbb{I}_2$ with $\sigma\in \{0.5,0.8,1\}$. We test different averaged number of tries $\widetilde{N}\in \{50, 100, 200, 500, 1000\}$. Thus, in the standard RW-MTM scheme, we set $N=\widetilde{N}$, whereas in the proposed mixture of MTM kernels in Eq. \eqref{Ducati}, we consider $M=3$ and  $N_1=1$, $N_2=\widetilde{N}$, $N_3=2\widetilde{N}-1$, so that we have always
$$
\widetilde{N}=\frac{N_1+N_2+N_3}{3}.
$$
 Therefore, the averaged computational cost is the same in both schemes, in terms of evaluations of the target distribution.   
The results, in terms of the expected number of iterations $E[\tau^*]$, are provided in Table \ref{table_RW_MTM}. First of all,   observe that, in general,  $E[\tau^*]$  grows if  the number of tries $N$ increases especially for the standard RW-MTM method (recall that for the standard RW- MTM scheme $N=\widetilde{N}$). The expected number of iterations $E[\tau^*]$ of the novel MTM technique with variable number of tries (introduced in Section \ref{PropSol1})  is always smaller than the corresponding value of the standard RW-MTM method. Namely, the novel scheme always outperforms the standard one, escaping from the region around ${\bf x}_0$ and reaching the modes of ${\bar \pi}({\bf x})$ more quickly, whereas the standard RW-MTM method remains stuck around ${\bf x}_0$ for several iterations, prejudicing its performance. Figures \ref{FigSimu} shows the improvement in the mixing with the proposed solution with respect to the standard RW-MTM technique. 

{\color{black}
Furthermore, the Mean Square Error (MSE) in the estimation of $E_{\pi}[{\bf X}]$ obtained by RW-MTM (and averaged over $500$ runs) is provided in Table \ref{table_RW_MTM2}. In this case, we set $\sigma=1$ and the initial state is chosen randomly ${\bf x}_0\sim \mathcal{U}([-6,6]\times [-6,6])$ (i.e., uniformly in the square $([-6,6]\times [-6,6]$), at each run. We can observe that the novel scheme provides always the smallest MSE confirming the robustness of the proposed solution.
}
\newline
{\bf I-MTM.} For the I-MTM scheme, we consider $N=2$ proposal pdfs and also $P=N=2$ number of tries (exactly as in the algorithm described in Table \ref{MTM_Table2}). Furthermore, the proposal pdfs are both Gaussians, specifically,  $q_n({\bf x})=\mathcal{N}({\bf x};{\bm \mu}_n,{\bm \Sigma})$, for $n=1,2$ and ${\bm \mu}_1=[-6,-6]^{\top}$, ${\bm \mu}_2=[0,0]^{\top}$ in the first configuration (denoted as {\bf Conf1}), and  ${\bm \mu}_1=[-6,-6]^{\top}$, ${\bm \mu}_2=[-1,-2]^{\top}$ in a second one (denoted as {\bf Conf2}). Thus, the second proposal pdf is always well-located, unlike the first one. The covariance matrix is the same for both proposals,  ${\bm \Sigma}=\sigma^2 \mathbb{I}_2$, and we test several values of $\sigma,$, i.e., $\sigma\in \{1.25,1.3,1.35,1.4\}$. As alternative scheme we consider the use of the deterministic mixture approach proposed in Section \ref{propSol2}. We  compute again the expected number of iterations $E[\tau^*]$ for reaching the modes starting from ${\bf x}_0=[-6,-6]^{\top}$ and set $T=4000$ as length of the chain, in this case. The results are provided in Table \ref{table_I_MTM}. We can observe that with the deterministic mixture approach the chain is able to jump easily to the regions of high probability of $\pi$, unlike with the standard I-MTM scheme. This occurs for every value of $\sigma$. With the standard I-MTM scheme the chain remains trapped around ${\bf x}_0$ for several iterations jeopardizing the performance of the algorithm (see also Table \ref{tableMSE_I_MTM2}).

{\color{black}
The  MSE values given in Table \ref{tableMSE_I_MTM2} (and averaged over $500$ runs) show that the improvement obtained by the novel scheme is even more evident than in the RW-MTM case. We have considered {\bf Conf2} and the initial state is chosen randomly ${\bf x}_0\sim \mathcal{U}([-6,6]\times [-6,6])$ at each run. 
} 
\begin{table}[!htb]
\caption{Expected number of iterations $E[\tau^*]$ required to escape from the region around ${\bf x}_0=[-6,-6]^{\top}$ with RW-MTM.}
\label{table_RW_MTM}
\footnotesize
\begin{center}
\begin{tabular}{|c|c|c|c|c|c|c|}
\hline
{\bf Scheme} & $\sigma$ &$ \widetilde{N}=50$ & $ \widetilde{N}=100$  & $ \widetilde{N}=200$ & $ \widetilde{N}=500$  & $ \widetilde{N}=1000$   \\
\hline
\hline
 standard   & \multirow{ 2}{*}{$0.5$} &  101.922 &  165.320  & 276.454 &  431.606 &  601.050 \\
 novel & & 67.237 & 72.349 &  81.253 &92.798  & 88.444  \\
\hline
\hline
 standard   & \multirow{ 2}{*}{$0.8$} & 205.299 &  367.358 &  612.442  & 1098.5 & 1363.1   \\
 novel &&  49.711 & 51.557  &49.405 & 49.706 &  56.145 \\
 \hline
 \hline
  standard   & \multirow{ 2}{*}{$1$} &  237.326 & 443.080  &709.808 & 784.644 & 699.614 \\
 novel & & 43.436 &41.236  & 33.906 &37.812 &  39.270 \\
  \hline
\end{tabular}
\end{center}
\end{table}%


\begin{table}[!htb]
{\color{black}
\caption{MSE in the estimation of $E_{\pi}[{\bf X}]$, obtained by RW-MTM, with $\sigma=1$ and ${\bf x}_0\sim \mathcal{U}([-6,6]\times [-6,6])$, i.e., randomly chosen at each run.
The standard and the novel scheme are test with different (fixed or averaged) number of tries $\widetilde{N}$. 
}
\label{table_RW_MTM2}
\footnotesize
\begin{center}
\begin{tabular}{|c|c|c|c|c|c|}
\hline
{\bf Scheme} &$ \widetilde{N}=50$ & $ \widetilde{N}=100$  & $ \widetilde{N}=200$ & $ \widetilde{N}=500$  & $ \widetilde{N}=1000$   \\
\hline
\hline
 standard   & 0.1702  & 0.1193   &  0.0892  & 0.0542   & 0.0266  \\
 novel   & 0.0533   &   0.0428 &  0.0329  &  0.0320 &  0.0228  \\
  \hline
\end{tabular}
\end{center}
}
\end{table}%


\begin{table}[!htb]
\caption{Expected number of iterations $E[\tau^*]$ required to escape from the region around ${\bf x}_0=[-6,-6]^{\top}$ with I-MTM.}
\label{table_I_MTM}
\footnotesize
\begin{center}
\begin{tabular}{|c|c|c|c|c|c|}
\hline
{\bf Scheme} & {\bf Conf} & $\sigma=1.25$  & $\sigma=1.3$  & $\sigma=1.35$  & $\sigma=1.4$   \\
\hline
\hline
 standard   & \multirow{ 2}{*}{1}    & 2967.6    & 1185.6 & 128.102   & 15.610  \\
 novel &  & 7.338 & 10.198 & 13.652  &  10.834      \\
\hline
\hline
 standard   & \multirow{ 2}{*}{2} & 3015.6  &  1212.9   & 139.816 & 20.548    \\
 novel & & 10.130 & 20.454   & 6.989    & 15.920 \\
 \hline
\end{tabular}
\end{center}
\end{table}


\begin{table}[!htb]
{\color{black}
\caption{MSE in the estimation of $E_{\pi}[{\bf X}]$, obtained by I-MTM, with {\bf Conf2}  and ${\bf x}_0\sim \mathcal{U}([-6,6]\times [-6,6])$, i.e.,  randomly chosen at each run.
}
\label{tableMSE_I_MTM2}
\footnotesize
\begin{center}
\begin{tabular}{|c|c|c|c|c|}
\hline
{\bf Scheme} & $\sigma=1.25$  & $\sigma=1.3$  & $\sigma=1.35$  & $\sigma=1.4$  \\
\hline
\hline
 standard   & 6.7943  & 6.4345   &  5.9183   &  5.5595      \\
 novel   & 0.7677   &  0.6987   &  0.3135   &     0.3055  \\
  \hline
\end{tabular}
\end{center}
}
\end{table}%

\section{Conclusions}
In this work, we have described different scenarios where MTM schemes have not the desired behavior, preventing the fast exploration of the state space.
These drawbacks cannot be solved simply increasing the computational effort, in terms of used number of tries. We have restricted the description of the problematic cases considering only the importance weights for the sake of simplicity, but the issues persist with other generic weight functions. Furthermore, we provide and discuss different solutions that solved the previously described problems, as also shown with numerical simulations. The proposed MTM schemes are in general more robust than the corresponding standard MTM techniques.

\section{Acknowledgements}
We would like to thank the Reviewers for their comments which have helped us to improve the  manuscript. This work has been supported by the Grant 2014/23160-6 of S\~ao Paulo Research Foundation (FAPESP) and by the Grant 305361/2013-3 of National Council for Scientific and Technological Development (CNPq).

\appendix
\section{Alternative weights in I-MTM}
\label{NoSol}

Other possible weight functions can be employed within MTM schemes without jeopardizing the ergodicity of the Markov chain. Let us consider the I-MTM scheme in Table \ref{MTM_Table2} using a generic weight function $w_n({\bf x})$, bounded and positive, i.e., $w_n({\bf x})> 0$, for all $n$. In this case, we have also to assume $\pi({\bf x})> 0$, for all $x\in \mathcal{X}$. 
 As shown in \citep{LucaJesse2,Pandolfi10}, the adequate probability for accepting the jump from ${\bf x}_{t-1}$ to ${\bf z}_j$ in this case is
\begin{equation}
\label{Estoca}
\alpha({\bf x}_{t-1},{\bf z}_j)=\min\left[1,\frac{\pi({\bf z}_j)q_j({\bf x}_{t-1})}{\pi({\bf x}_{t-1})q_j({\bf z}_j)}\frac{W_X}{W_Z}\right],
\end{equation}
where
$$
W_Z=\frac{w_j({\bf z}_j)}{\sum_{n=1}^N w_n({\bf z}_n)}, \quad\quad W_X=\frac{w_j({\bf x}_{t-1})}{\left[\sum_{n=1}^N w_n({\bf z}_n)\right]-w_j({\bf z}_j)+w_j({\bf x}_{t-1})}.
$$
If the chosen weights are the importance weights, $w_n({\bf x})=\frac{\pi({\bf x})}{q_n({\bf x})}$, then Eq. \eqref{Estoca} coincides with Eq. \eqref{AlfaMTM2}. Moreover, note that, in any case, $0 \leq W_Z\leq 1$ and $0 \leq W_X\leq 1$. As explained in Section \ref{OtroProbl}, in general, it often occurs that  $q_j({\bf z}_j)>q_j({\bf x}_{t-1})$ since ${\bf z}_j \sim q_j({\bf z})$ whereas ${\bf x}_{t-1}$ has been generated from a generic $q_k$ with $k\in \{1,\ldots,N\}$. Thus, $\frac{\pi({\bf z}_j)q_j({\bf x}_{t-1})}{\pi({\bf x}_{t-1})q_j({\bf z}_j)}$ tends to be close to zero and as consequence often $\alpha\approx 0$, regardless of the choice of the weight functions. Observe that if we employ the set of proposal pdfs $q_j ({\bf x})$'s as a mixture $\psi({\bf x})=\frac{1}{N}\sum_{n=1}^N q_n({\bf x})$ as suggested in Section \ref{propSol2}, the problem is solved also in this case.


\begin{thebibliography}{25}
\providecommand{\natexlab}[1]{#1}
\providecommand{\url}[1]{\texttt{#1}}
\expandafter\ifx\csname urlstyle\endcsname\relax
  \providecommand{\doi}[1]{doi: #1}\else
  \providecommand{\doi}{doi: \begingroup \urlstyle{rm}\Url}\fi

\bibitem[Robert and Casella(2004)]{Robert04}
C.~P. Robert and G.~Casella.
\newblock \emph{{M}onte {C}arlo Statistical Methods}.
\newblock Springer, 2004.

\bibitem[Liu(2004)]{Liu04b}
J.~S. Liu.
\newblock \emph{{M}onte {C}arlo Strategies in Scientific Computing}.
\newblock Springer, 2004.

\bibitem[Liang et~al.(2010)Liang, Liu, and Caroll]{Liang10}
F.~Liang, C.~Liu, and R.~Caroll.
\newblock \emph{Advanced {M}arkov {C}hain {M}onte {C}arlo Methods: Learning
  from Past Samples}.
\newblock Wiley Series in Computational Statistics, England, 2010.

\bibitem[Liu et~al.(2000)Liu, Liang, and Wong]{Liu00}
J.~S. Liu, F.~Liang, and W.~H. Wong.
\newblock The multiple-try method and local optimization in metropolis
  sampling.
\newblock \emph{Journal of the American Statistical Association}, 95\penalty0
  (449):\penalty0 121--134, March 2000.

\bibitem[Metropolis et~al.(1953)Metropolis, Rosenbluth, Rosenbluth, Teller, and
  Teller]{Metropolis53}
N.~Metropolis, A.~Rosenbluth, M.~Rosenbluth, A.~Teller, and E.~Teller.
\newblock Equations of state calculations by fast computing machines.
\newblock \emph{Journal of Chemical Physics}, 21:\penalty0 1087--1091, 1953.

\bibitem[Hastings(1970)]{Hastings70}
W.~K. Hastings.
\newblock {M}onte {C}arlo sampling methods using {M}arkov chains and their
  applications.
\newblock \emph{Biometrika}, 57\penalty0 (1):\penalty0 97--109, 1970.

\bibitem[Frenkel and Smit(1996)]{Frenkel96}
D.~Frenkel and B.~Smit.
\newblock Understanding molecular simulation: from algorithms to applications.
\newblock \emph{Academic Press, San Diego}, 1996.

\bibitem[Qin and Liu(2001)]{Qin01}
Z.~S. Qin and J.~S. Liu.
\newblock {M}ulti-{P}oint {M}etropolis method with application to hybrid
  {M}onte {C}arlo.
\newblock \emph{Journal of Computational Physics}, 172:\penalty0 827--840,
  2001.

\bibitem[Casarin et~al.(2013)Casarin, Craiu, and Leisen]{Casarin2011}
R.~Casarin, R.~Craiu, and F.~Leisen.
\newblock Interacting multiple try algorithms with different proposal
  distributions.
\newblock \emph{Statistics and Computing}, 23\penalty0 (2):\penalty0 185--200,
  2013.

\bibitem[Pandolfi et~al.(2010)Pandolfi, Bartolucci, and Friel]{Pandolfi10}
Silvia Pandolfi, Francesco Bartolucci, and Nial Friel.
\newblock A generalization of the {M}ultiple-try {M}etropolis algorithm for
  {B}ayesian estimation and model selection.
\newblock \emph{Journal of Machine Learning Research (Workshop and Conference
  Proceedings Volume 9: AISTATS 2010)}, 9:\penalty0 581--588, 2010.

\bibitem[Martino et~al.(2012)Martino, Olmo, and Read]{LucaJesse1}
L.~Martino, V.~P.~Del Olmo, and J.~Read.
\newblock A multi-point {M}etropolis scheme with generic weight functions.
\newblock \emph{Statistics \& Probability Letters}, 82\penalty0 (7):\penalty0
  1445--1453, July 2012.

\bibitem[Craiu and Lemieux(2007)]{Craiu07}
R.~V. Craiu and C.~Lemieux.
\newblock Acceleration of the {M}ultiple {T}ry {M}etropolis algorithm using
  antithetic and stratified sampling.
\newblock \emph{Statistics and Computing}, 17\penalty0 (2):\penalty0 109--120,
  June 2007.

\bibitem[B{\'e}dard et~al.(2012)B{\'e}dard, Douc, and Mouline]{Bedard12}
M.~B{\'e}dard, R.~Douc, and E.~Mouline.
\newblock Scaling analysis of multiple-try {MCMC} methods.
\newblock \emph{Stochastic Processes and their Applications}, 122:\penalty0
  758--786, 2012.

\bibitem[Martino and Read(2013)]{LucaJesse2}
L.~Martino and J.~Read.
\newblock On the flexibility of the design of multiple try {M}etropolis
  schemes.
\newblock \emph{Computational Statistics}, 28\penalty0 (6):\penalty0
  2797--2823, December 2013.

\bibitem[Martino et~al.(2014)Martino, Leisen, and Corander]{MartinoMTM2014}
L.~Martino, F.~Leisen, and J.~Corander.
\newblock On {M}ultiple {T}ry schemes and the {P}article
  {M}etropolis-{H}astings algorithm.
\newblock \emph{viXra:1409.0051}, 2014.

\bibitem[Martino et~al.(2015{\natexlab{a}})Martino, Elvira, Luengo, Corander,
  and Louzada]{O-MCMC}
L.~Martino, V.~Elvira, D.~Luengo, J.~Corander, and F.~Louzada.
\newblock Orthogonal parallel {MCMC} methods for sampling and optimization.
\newblock \emph{arXiv:1507.08577}, 2015{\natexlab{a}}.

\bibitem[Veach and Guibas(1995)]{Veach95}
E.~Veach and L.~Guibas.
\newblock Optimally combining sampling techniques for {M}onte {C}arlo
  rendering.
\newblock \emph{In SIGGRAPH 1995 Proceedings}, pages 419--428, 1995.

\bibitem[Owen and Zhou(2000)]{Owen00}
A.~Owen and Y.~Zhou.
\newblock Safe and effective importance sampling.
\newblock \emph{Journal of the American Statistical Association}, 95\penalty0
  (449):\penalty0 135--143, 2000.

\bibitem[Elvira et~al.(2015{\natexlab{a}})Elvira, Martino, Luengo, and
  Bugallo]{LetterVictor}
V.~Elvira, L.~Martino, D.~Luengo, and M.~Bugallo.
\newblock Efficient multiple importance sampling estimators.
\newblock \emph{IEEE Signal Processing Letters}, 22\penalty0 (10):\penalty0
  1757--1761, 2015{\natexlab{a}}.

\bibitem[Elvira et~al.(2015{\natexlab{b}})Elvira, Martino, Luengo, and
  Bugallo]{TutorialMIS}
V.~Elvira, L.~Martino, D.~Luengo, and M.~F. Bugallo.
\newblock Generalized multiple importance sampling.
\newblock \emph{arXiv:1511.03095}, 2015{\natexlab{b}}.

\bibitem[Cornuet et~al.(2012)Cornuet, Marin, Mira, and Robert]{CORNUET12}
J.~M. Cornuet, J.~M. Marin, A.~Mira, and C.~P. Robert.
\newblock Adaptive multiple importance sampling.
\newblock \emph{Scandinavian Journal of Statistics}, 39\penalty0 (4):\penalty0
  798--812, December 2012.

\bibitem[Martino et~al.(2015{\natexlab{b}})Martino, Elvira, Luengo, and
  Corander]{APIS15}
L.~Martino, V.~Elvira, D.~Luengo, and J.~Corander.
\newblock An adaptive population importance sampler: Learning from the
  uncertanity.
\newblock \emph{IEEE Transactions on Signal Processing}, 63\penalty0
  (16):\penalty0 4422--4437, 2015{\natexlab{b}}.

\bibitem[Martino et~al.(2015{\natexlab{c}})Martino, Elvira, Luengo, and
  Corander]{LAIS2015}
L.~Martino, V.~Elvira, D.~Luengo, and J.~Corander.
\newblock Layered adaptive importance sampling.
\newblock \emph{arXiv:1505.04732}, 2015{\natexlab{c}}.

\bibitem[Ali et~al.(2007)Ali, Yao, Collier, Taylor, Blumstein, and
  Girod]{Ali07}
A.~M. Ali, K.~Yao, T.~C. Collier, E.~Taylor, D.~Blumstein, and L.~Girod.
\newblock An empirical study of collaborative acoustic source localization.
\newblock \emph{Proc. Information Processing in Sensor Networks (IPSN07),
  Boston}, April 2007.

\bibitem[Fitzgerald(2001)]{Fitzgerald01}
W.~J. Fitzgerald.
\newblock {M}arkov chain {M}onte {C}arlo methods with applications to signal
  processing.
\newblock \emph{Signal Processing}, 81\penalty0 (1):\penalty0 3--18, January
  2001.

\end{thebibliography}
\end{document}